\documentstyle[prd,aps,epsf]{revtex}

\newcommand{\STRUT}{\rule{0in}{2.5ex}}

\def\mps{m_{\rm PS}}
\def\mv{m_{\rm V}}

\def\simgt{\,\rlap{\lower 3.5 pt\hbox{$\mathchar \sim$}}\raise 1pt \hbox {$>$}\,}
\def\simlt{\,\rlap{\lower 3.5 pt\hbox{$\mathchar \sim$}}\raise 1pt \hbox {$<$}\,}

\newcommand{\plaq}{
   \put(-.5,-.5){\line(1,0){1}}
   \put(.5,-.5){\line(0,1){1}}
   \put(.5,.5){\line(-1,0){1}}
   \put(-.5,.5){\line(0,-1){1}}
}

\newcommand{\recly}{
   \put(-1.,-.5){\line(1,0){2}}
   \put(-1.,.5){\line(1,0){2}}
   \put(-1.,-.5){\line(0,1){1}}
   \put(1.,-.5){\line(0,1){1}}
}

\newcommand{\recst}{
   \put(-.5,-1.){\line(1,0){1}}
   \put(-.5,1.){\line(1,0){1}}
   \put(-.5,-1.){\line(0,1){2}}
   \put(.5,-1.){\line(0,1){2}}
}

\newcommand{\clover}{\setlength{\unitlength}{.4cm}\raisebox{-.4cm}{
   \begin{picture}(2.7,2.3)(-1.15,-1.15)
   \multiput(-1.15,-1.15)(1.15,1.15){2}{\begin{picture}(1.15,1.15)(-.575,-.575)
   \plaq\end{picture}}
   \multiput(-1.15,0)(1.15,-1.15){2}{\begin{picture}(1.15,1.15)(-.575,-.575)
   \plaq\end{picture}}
   \put(0,0){\circle*{.3}}
   \end{picture}}}

\newcommand{\clrecly}{\setlength{\unitlength}{.4cm}\raisebox{-.4cm}
{  \begin{picture}(2.6,2.6)(-1.55,-1.15)
   \multiput(-1.15,-0)(2.15,0.){2}{\begin{picture}(1.15,1.15)(-.575,-.575)
   \recly\end{picture}}
   \multiput(-1.15,-1.15)(2.15,0.){2}{\begin{picture}(1.15,1.15)(-.575,-.575)
   \recly\end{picture}}
   \put(.5,0){\circle*{.3}}
   \end{picture}}}

\newcommand{\clrecst}{\setlength{\unitlength}{.4cm}\raisebox{-.4cm}
{   \begin{picture}(2.6,2.6)(-1.55,-1.55)
   \multiput(-2.15,0)(1.15,0){2}{\begin{picture}(1.15,1.15)(-.575,-.575)
   \recst\end{picture}}  
   \multiput(-2.15,-2.15)(1.15,0){2}{\begin{picture}(1.15,1.15)(-.575,-.575)
   \recst\end{picture}}  
   \put(-1.0,-.5){\circle*{.3}}
   \end{picture}}}

\begin{document}

\title{
%
%
\begin{flushright}
\normalsize
UTCCP-P-107          \\
\end{flushright}
%
%
Topological Susceptibility in Lattice QCD with Two Flavors of Dynamical 
Quarks}  

\author{A.~Ali Khan,$^1$\thanks{address till 31 August, 2000}
        S.~Aoki,$^2$
        R.~Burkhalter,$^{1,2}$
        S.~Ejiri,$^1$\thanks{present address :
            Department of Physics, University of Wales, 
            Swansea SA2 8PP, U.K.}
        M.~Fukugita,$^3$
        S.~Hashimoto,$^4$
        N.~Ishizuka,$^{1,2}$
        Y.~Iwasaki,$^{1,2}$
        K.~Kanaya,$^{2}$
        T.~Kaneko,$^4$
        Y.~Kuramashi,$^4$
        T.~Manke,$^1$\thanks{present address :
            Physics Department, Columbia University, 
            New York, NY 10027, U. S. A.}
        K.~Nagai,$^1$\thanks{present address :
            CERN, Theory Division, CH--1211 Geneva 23, Switzerland}
        M.~Okawa,$^4$
        H.P.~Shanahan,$^1$\thanks{present address :
            Department of Biochemistry and Molecular
            Biology, University College London, London, England, U.K.}
        A.~Ukawa,$^{1,2}$ and
        T.~Yoshi\'e$^{1,2}$ \\[2mm]
        (CP-PACS Collaboration)
}

\address{
$^1$Center for Computational Physics,
University of Tsukuba, Tsukuba, Ibaraki 305--8577, Japan \\
$^2$Institute of Physics, University of
Tsukuba, Tsukuba, Ibaraki 305--8571, Japan \\
$^3$Institute for Cosmic Ray Research,
University of Tokyo, Kashiwa 277--8582, Japan \\
$^4$High Energy Accelerator Research Organization
(KEK), Tsukuba, Ibaraki 305--0801, Japan
}

\date{June 16, 2001}

\maketitle

\begin{abstract}

We present a study of the topological susceptibility in lattice QCD with
two degenerate flavors of dynamical quarks. The topological charge is
measured on gauge configurations generated with a renormalization group
improved gauge action and a mean field improved clover quark action at
three values of $\beta=6/g^2$, corresponding to lattice spacings of $a
\approx 0.22$, 0.16 and 0.11~fm, with four sea quark masses at each $\beta$. 
The study is supplemented by simulations of pure SU(3) gauge theory with the
same gauge action at 5 values of $\beta$ with lattice spacings
0.09~fm$\simlt a \simlt$0.27~fm. We employ a field theoretic definition of
the topological charge together with cooling. For the topological
susceptibility in the continuum limit of pure SU(3) gauge theory we obtain
$\chi_t^{1/4} = 197^{+13}_{-16}$~MeV where the error shows statistical and
systematic ones added in quadrature.  In full QCD $\chi_t$ at heavy sea
quark masses is consistent with that of pure SU(3) gauge theory. A decrease of
$\chi_t$ toward light quark masses, as predicted by the anomalous
Ward-Takahashi identity for U(1) chiral symmetry, becomes clearer for
smaller lattice spacings.  The cross-over in the behavior of $\chi_t$ 
from heavy to light sea quark masses is discussed.

\end{abstract}


\newpage


\section{Introduction}
\label{sec:intro}

The topological structure of gauge field fluctuations, in particular
instantons, has been invoked to explain several important low energy
properties of QCD including the breaking of axial U(1) symmetry and the
large mass of the $\eta'$ meson. Numerical simulations on a space-time
lattice provide a non-perturbative tool for the study of these phenomena
beyond semiclassical approximations.

Lattice studies of the topological susceptibility $\chi_t$ as a measure of
these fluctuations have been mostly carried out for pure gauge
theory without the presence of dynamical
fermions~\cite{pure-review}. Recent determinations by various groups using
different methods have led to a consistent value in SU(3) gauge theory of
$\chi_t^{1/4} = 200 \pm 18$~MeV~\cite{pure-review}.

Sea quark effects on the topological susceptibility have been much less
studied, although dynamical quarks are expected to have a strong influence
on $\chi_t$ leading to a complete suppression for massless quarks. From the
anomalous Ward-Takahashi identity for U(1) chiral symmetry, the topological
susceptibility is predicted~\cite{Crewther,DiVecchia,LS} to obey for small
quark masses in the chirally broken phase,
\begin{equation}
\chi_t = \frac{\Sigma}{N_f}m_q + O(m_q^2), \;\; 
\Sigma=-\lim_{m_q\rightarrow0}
\lim_{V\rightarrow\infty}\langle\overline{\psi}\psi\rangle.
\label{eq:chismallmass} 
\end{equation}
It is an interesting question to investigate whether lattice data confirm a 
suppression consistent with Eq.~(\ref{eq:chismallmass}).

Pioneering attempts to calculate $\chi_t$ in full
QCD~\cite{Gausterer,Bitar,Kuramashi} were restricted to small statistics
and were plagued by long autocorrelation times. Progress in the simulation
of full QCD, as well as increase of available computer power in recent
years, has enabled this question to be readdressed with a higher accuracy. 
A number of pieces of work have been reported
recently~\cite{EtaTopo,Hart,UKQCD,SESAM2,Hasenfratz,pisa} coming to
different conclusions whether the topological susceptibility is consistent
with the prediction of Eq.~(\ref{eq:chismallmass}). A common shortcoming in
Refs.~\cite{EtaTopo,Hart,UKQCD,SESAM2,Hasenfratz} is that they have been
made at only one lattice spacing. Ref.~\cite{pisa}, on the other hand, used
only one bare quark mass $am_q$ at each coupling constant $\beta$.

In this article we attempt to improve on this status by calculating the
topological susceptibility in full QCD with two flavors of dynamical quarks
at four sea quark masses at each of three gauge couplings. We perform
calculations on configurations of the CP-PACS full QCD project~\cite{full}. 
These have been generated on the CP-PACS parallel
computer~\cite{CPPACS-machine} using a renormalization group (RG) improved
gauge action~\cite{RGIA} and a mean field improved Sheikholeslami-Wohlert
clover quark action~\cite{clover}.  The efficacy of this choice of action
over the standard action has been demonstrated in Ref.~\cite{comparative}
by examining both the rotational symmetry of the static quark potential and
the scaling behavior of light hadron mass ratios.

Preliminary results for the topological susceptibility based on a
first analysis at our intermediate lattice spacing have been published in
Ref.~\cite{EtaTopo}. In this article we present the final analysis and
results at all gauge couplings. 

The identification of dynamical quark effects requires a comparison 
with pure SU(3) gauge theory where sea quarks are absent. We therefore
supplement our study of topology in full QCD by a set of simulations of
SU(3) gauge theory with the same RG-improved gluon action at a similar
range of lattice spacings. 

The outline of this article is as follows. In Sec.~\ref{sec:simulations} we
give details on numerical simulations and measurements of the topological
charge. Results for the topological susceptibility are presented in
Sec.~\ref{sec:suscept} where we discuss the continuum extrapolation in pure
gauge theory, as well as the quark mass dependence in full QCD. Conclusions
are summarized in Sec.~\ref{sec:conclusions}.


\section{Computational Details}
\label{sec:simulations}

\subsection{Gauge configurations}

Gauge configurations incorporating two degenerate flavors of dynamical
quarks have been generated by the CP-PACS full QCD project. For gluons we
employed an RG-improved action~\cite{RGIA} of the form,
\begin{equation}
S_{\rm RG} = 
\frac{\beta}{6} \left\{\;3.648\sum_{x,\mu <\nu} W_{\mu\nu}^{1\times 1}(x) 
-0.331\sum_{x,\mu ,\nu} W_{\mu\nu}^{1\times 2}(x)\right\},
\label{eq:RGAct} 
\end{equation}
where $W^{1\times 1}$ and $W^{1\times 2}$ are the plaquette and rectangular
Wilson loop. For the quark part we adopted the clover quark
action~\cite{clover} with a mean field improved clover coefficient $c_{\rm
SW}=P^{-3/4}$, and the plaquette $P$ calculated in perturbation
theory at one loop $P=1-0.8412\beta^{-1}$. This choice is based on the
observation that measured values of the plaquette $\langle P \rangle$ are
well approximated by the one-loop estimate~\cite{full} and that $c_{\rm
SW}$ determined in this way is close to its one-loop value~\cite{aoki}.

Three sets of gauge configurations have been generated at bare gauge
couplings $\beta=1.8$, 1.95 and 2.1 corresponding to the lattice spacings
$a\approx 0.22$, 0.16 and 0.11~fm. Lattices of size $L^3\times T =
12^3\times 24$, $16^3\times 32$ and $24^3\times 48$ have been used, for
which the physical lattice size remains approximately constant at
$La\approx 2.5$~fm. At each $\beta$, runs are carried out at four values of
the hopping parameter $\kappa$ chosen such that the mass ratio of
pseudoscalar to vector mesons takes $\mps/\mv \approx 0.8$, 0.75, 0.7 and
0.6.

In Table~\ref{tab:overview} we give an overview of the parameters and
statistics of the full QCD runs. Technical details concerning the
configuration generation with the Hybrid Monte Carlo (HMC) algorithm and
results for the light hadron spectrum are presented in
Ref.~\cite{full}. Runs were made with a length of 4000--7000 HMC
unit-trajectories per sea quark mass. Topology measurements are made on
configurations separated by 10 HMC trajectories at $\beta=1.8$ and 1.95 and
by 5 trajectories at $\beta=2.1$. The number of measurements $N_{\rm Meas}$
and the separations $N_{\rm Skip}$ are listed for each run in
Table~\ref{tab:overview}.

We supplement the study of topology in full QCD by simulations of pure
SU(3) gauge theory with the RG-improved action of
Eq.~(\ref{eq:RGAct}). Configurations are generated at 5 values of $\beta$
with lattice spacings 0.09~fm$\simlt a \simlt$0.27~fm as listed in
Table~\ref{tab:overview-pure}. For the three larger gauge couplings
lattices of size $8^4$, $12^4$ and $16^4$ are used so that the physical
lattice size remains approximately constant at $La\approx 1.5$~fm. While
this is smaller than the sizes in the full QCD runs, it has been a standard
size employed in recent studies of topology in SU(3) gauge
theory~\cite{pisa-SU(3),ImpCool-SU(3),UKQCD-SU3,boulder-SU(3)}. It has also
been shown~\cite{UKQCD-SU3} that the instanton size distribution does not
suffer from significant finite volume effects on a lattice of this
size. For the two smaller gauge couplings we keep lattices of size
$8^4$. Simulations are carried out with a combination of the
pseudo-heat-bath algorithm and the over-relaxation algorithm mixed in a
ratio $1:4$. For each $\beta$ we create 500--2000 independent
configurations separated by 100 iterations.


\subsection{Topological charge operator}

The topological charge density in the continuum is defined by
\begin{equation}
Q(x) = \frac{1}{32\pi^2} \epsilon_{\mu\nu\rho\sigma} \;
{\rm Tr} \left ( F_{\mu\nu}(x)F_{\rho\sigma}(x) \right ),
\end{equation}
and the total topological charge $Q$ is an integer defined by the
integrated form 
\begin{equation}
Q = \int d^4x \; Q(x).
\end{equation}

On the lattice we use the field theoretic transcription of this operator
which has the standard form
\begin{equation}
Q_{\rm st}  =  \sum_{x_n} Q_L^P(x_n), 
\label{eq:Qstand} 
\end{equation}
with the lattice charge density defined by
\begin{equation}
Q_L^P(x_n) = \frac{1}{32\pi^2} \epsilon_{\mu\nu\rho\sigma} \;
{\rm Tr} \left ( C^P_{\mu\nu}(x)C^P_{\rho\sigma}(x) \right ).
\end{equation}
In this expression the field strength on the lattice is defined through the 
clover leaf
\begin{equation}
C^P_{\mu\nu} = \frac{1}{4} \; {\rm Im} \left ( \clover \right ).
\end{equation}

An improved charge operator can be constructed by additionally calculating
a rectangular clover leaf made out of $1\times 2$ Wilson loops,
\begin{equation}
C^R_{\mu\nu} = \frac{1}{8} \; {\rm Im} \left ( \clrecly \hspace{0.8cm} 
+ \ \ \clrecst \right ),
\end{equation}
and combine them to the charge density
\begin{equation}
Q_L^R(x_n) = \frac{2}{32\pi^2} \epsilon_{\mu\nu\rho\sigma} \;
{\rm Tr} \left ( C^R_{\mu\nu}(x)C^R_{\rho\sigma}(x) \right ).
\end{equation}
The improved global charge is then defined through
\begin{equation}
Q_{\rm imp} = \sum_{x_n} \left \{ c_0 Q_L^P(x_n) + c_1 Q_L^R(x_n) \right \} .
\label{eq:Qimp}
\end{equation}

The standard charge operator of Eq.~(\ref{eq:Qstand}) has $O(a^2)$
discretization errors. With the choice $c_0=5/3$ and
$c_1=-1/12$~\cite{Weisz,LW} in Eq.~(\ref{eq:Qimp}) the leading order $a^2$
terms are removed for classical instanton configurations and discretization
errors become $O(a^4)$.


\subsection{Cooling}

The topological charge operators of Eq.~(\ref{eq:Qstand}) or
(\ref{eq:Qimp}) are dominated by local fluctuations of gauge fields when
measured on thermalized lattice configurations and their value is generally
noninteger. The cooling method~\cite{cooling} removes the ultraviolet
fluctuations by minimizing the action locally while not significantly
disturbing the underlying long-range topological structure.

In full QCD one might consider cooling with the full action including the
fermionic part. We refrain from this because it would lead to solutions of
the classical equations of motion of the effective action, obtained by
integrating out fermion fields~\cite{Kogut}. These are different from
instantons which are solutions of the classical equations of motion of the
gauge action only. Moreover, cooling would become a non-local process.

In principle any lattice discretization of the continuum gauge action can
be used for smoothing gauge configurations by cooling. However, lattice
actions generally do not have scale invariant instanton solutions. The
standard Wilson plaquette action discretization of a continuum instanton
solution with radius $\rho$, for example, behaves for $a \ll \rho \ll L$
as~\cite{OvImpCool} 
\begin{equation}
S_{\rm plaq} = S_{\rm cont} \left \{ 1 - \frac{1}{5} (a/\rho)^2 
                     + O \left ((a/\rho)^4 \right ) \right \}.
\end{equation}
Under cooling with the plaquette action, instantons therefore shrink and
disappear when the cooling is applied too long. To improve on this we use
for cooling a gluon action of the generic form 
\begin{equation}
S_{\rm cool} = 
\left\{\;c_0\sum_{x,\mu <\nu} W_{\mu\nu}^{1\times 1}(x) 
+c_1\sum_{x,\mu ,\nu} W_{\mu\nu}^{1\times 2}(x)\right\},
\label{eq:CoolAct} 
\end{equation}
where the coefficients $c_0$ and $c_1$ satisfy the normalization condition
$c_0 + 8c_1 = 1$. We employ the two choices
\begin{equation}
c_0 = 5/3,\;   c_1 = -1/12  \;\;\;  \mbox{LW action,}  
\label{eq:LW}
\end{equation}
and
\begin{equation}
c_0 = 3.648,\; c_1 = -0.331 \;\;\; \mbox{RG-improved action.} 
\label{eq:RG}
\end{equation}

The tree-level improved Symanzik action by L\"uscher and
Weisz\cite{Weisz,LW} of Eq.~(\ref{eq:LW}) has reduced breaking of instanton
scale invariance given by~\cite{OvImpCool},
\begin{equation}
S_{\rm LW} = S_{\rm cont} \left \{ 1 - \frac{17}{210} (a/\rho)^4 
                     + O \left ((a/\rho)^6 \right ) \right \},
\end{equation}
while still not admitting stable instantons under cooling. 
For the RG-improved action of Eq.~(\ref{eq:RG}) the sign 
of the leading order term is changed~\cite{OvImpCool}, 
\begin{equation}
S_{\rm RG} = S_{\rm cont} \left \{ 1 + \frac{2.972}{5} (a/\rho)^2 
                     + O \left ((a/\rho)^4 \right ) \right \}.
\end{equation}
The flip of the sign leads to a local minimum of the
action where stable lattice instantons can exist~\cite{IIY}.

Cooling with the RG-improved 
action or the LW action can lead to different values of
the topological charge since instantons with a radius of the order of the
lattice spacing can be either destroyed or stablized. The ambiguity is only
expected to vanish when the lattice is fine enough. We test this explicitly
by using both actions for cooling and treat differences as a systematic
error of the cooling method.

A cooling step consists of the minimization of the local action for
three SU(2) subgroups at every link of the lattice using the
pseudo-heat-bath algorithm with $\beta=\infty$. We have made 50 cooling
steps for every configuration, measuring the topological charge after each
step.  

\begin{figure*}[p]
\vspace{5mm}
\centerline{
              \epsfxsize=8cm \epsfbox{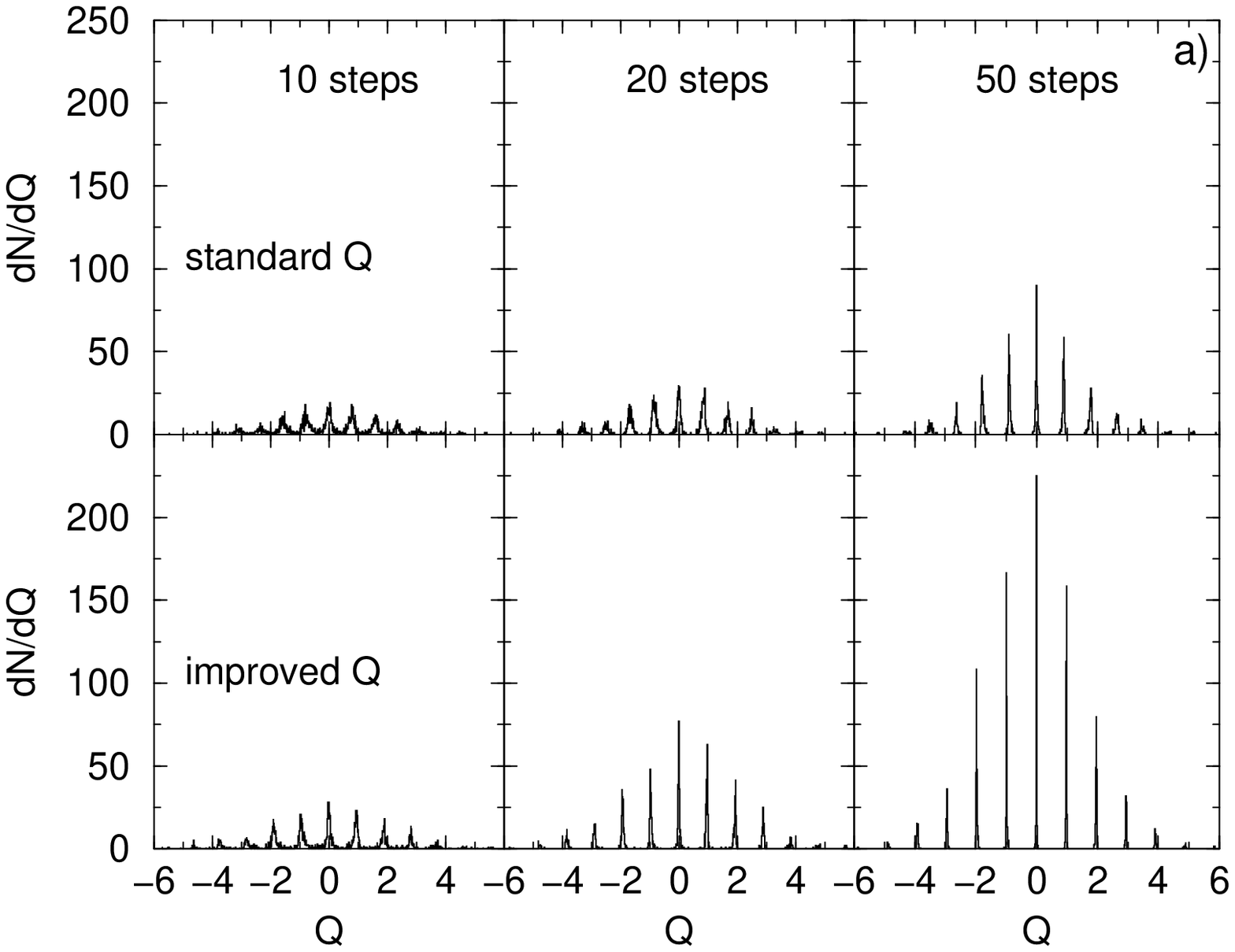}
\hspace{5mm}  \epsfxsize=8cm \epsfbox{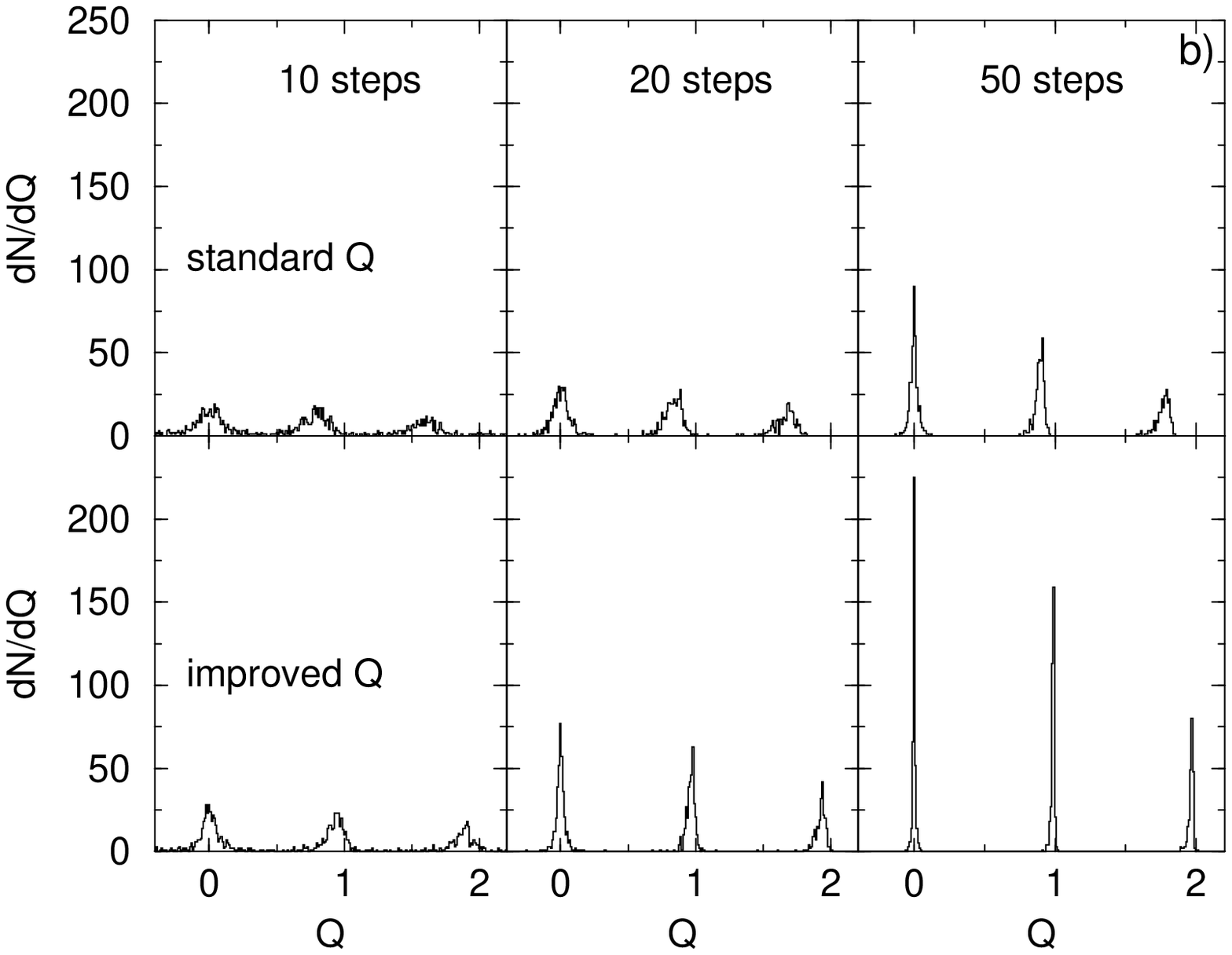}
}
\caption{Topological charge distribution at $\beta=2.227$ after various
numbers of cooling steps with the RG-improved 
action and for two definitions of the
topological charge. Figure a) shows the whole distribution while Fig.~b) is 
an enlargement of the first three peaks.} 
\label{fig:Integer}
\end{figure*}

We have investigated the deviations from integer topological charge as a
function of the number of cooling steps, the topological charge operator,
and the coupling constant for our simulations of pure SU(3) gauge
theory. In Fig.~\ref{fig:Integer} we show the distribution of the
topological charge at the intermediate gauge coupling of $\beta=2.227$.
The distribution is peaked at quantized but noninteger values of $Q$. The
peaks are already well separated after 10 cooling steps and the widths of
peaks further decrease with increasing number of cooling steps. At the same
number of cooling steps, peaks are narrower for the improved charge
operator $Q_{\rm imp}$ than for the naive form $Q_{\rm st}$. Centers of
peaks are located below integer values. Cooling and improvement of charge
operator move them closer to integers. In Table~\ref{tab:DevInt} we list
the ratio between center of peaks and integer charge, found to be
independent of $Q$, for all gauge couplings and after 10, 20, or 50 cooling
steps with the RG-improved action. The ratio moves closer to unity with
increasing gauge coupling, increasing number of cooling steps and when the
charge operator is improved, showing that the difference from integer is a
finite lattice spacing effect. After 20 cooling steps, centers of peaks of
$Q_{\rm imp}$ do not differ from integer by more than 6\% even at the
coarsest lattice spacing. Because of its superiority we only use $Q_{\rm
imp}$, rounded to nearest integer, in the following.

\begin{figure*}[p]
\vspace{5mm}
\centerline{
\epsfxsize=9cm \epsfbox{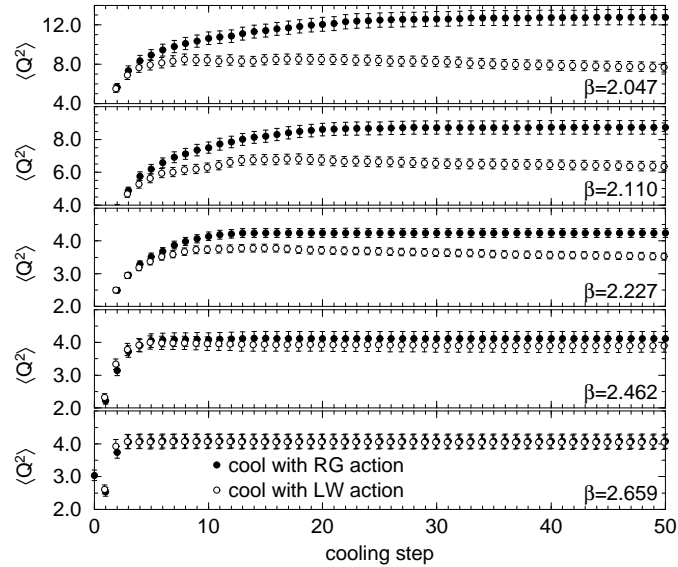}
}
\caption{Expectation value of the topological charge squared as a function of 
the number of cooling steps for two different cooling actions.}
\label{fig:QintsqCool}
\end{figure*}

In Fig.~\ref{fig:QintsqCool} we plot $\langle Q^2 \rangle$ measured in pure 
SU(3) gauge theory as a function of the number of cooling steps for the two
cooling actions. Cooling with the two actions leads to quite different
values of $\langle Q^2 \rangle$ at coarser lattice spacings. The difference 
decreases with increasing coupling constant and almost vanishes on the
finest lattice.

We quantify the difference between cooling with the two actions by
calculating the linear correlation coefficient
\begin{equation}
r = \frac{\left\langle 
\left(Q_{\rm imp}^{\rm RG} - \overline{Q_{\rm imp}^{\rm RG}}\right) 
\left(Q_{\rm imp}^{\rm LW} - \overline{Q_{\rm imp}^{\rm LW}}\right) 
\right\rangle}
{\sqrt{\left\langle 
\left(Q_{\rm imp}^{\rm RG} - \overline{Q_{\rm imp}^{\rm RG}}\right)^2 
\right\rangle
\left\langle 
\left(Q_{\rm imp}^{\rm LW} - \overline{Q_{\rm imp}^{\rm LW}}\right)^2 
\right\rangle}},
\label{eq:corrcoeff}
\end{equation}
after 10, 20, or 50 cooling steps. For the evaluation of
Eq.~(\ref{eq:corrcoeff}) we substitute charges before rounding to integers.
Values of $r$ are listed in Table~\ref{tab:corrRGLW}. The correlation
between topological charge after cooling with the RG-improved or the LW action
decreases with increasing number of cooling steps. Even at the coarsest
lattice spacing and after 50 cooling steps, however, there is a strong
correlation with $r=0.84$. With decreasing lattice spacing $r$ approaches
unity and charges are highly correlated on the finest lattice. These
features agree with our naive expectations. 

Since $\langle Q^2 \rangle$ has an approximate plateau after 20 cooling
steps we use this as central value. $\langle Q^2 \rangle$ is listed for
pure SU(3) gauge theory in Table~\ref{tab:suscept-pure} and for full QCD in
Table~\ref{tab:suscept}. The first quoted error is statistical. The
second error expresses the uncertainty of choosing the number of cooling
steps by taking the largest difference between $\langle Q^2 \rangle$ after
20 cooling steps and after more cooling steps up to 50.


\subsection{Full QCD time histories}

Decorrelation of topology is an important issue in the simulation of full
QCD since the topological charge is one of the quantities which is expected 
to have the longest autocorrelation with the HMC algorithm. In simulations
with the Kogut-Susskind quark action it was found that topological modes
have a very long autocorrelation time~\cite{Kuramashi,Boyd}.

\begin{figure*}[p]
\vspace{5mm}
\centerline{
\epsfxsize=10cm \epsfbox{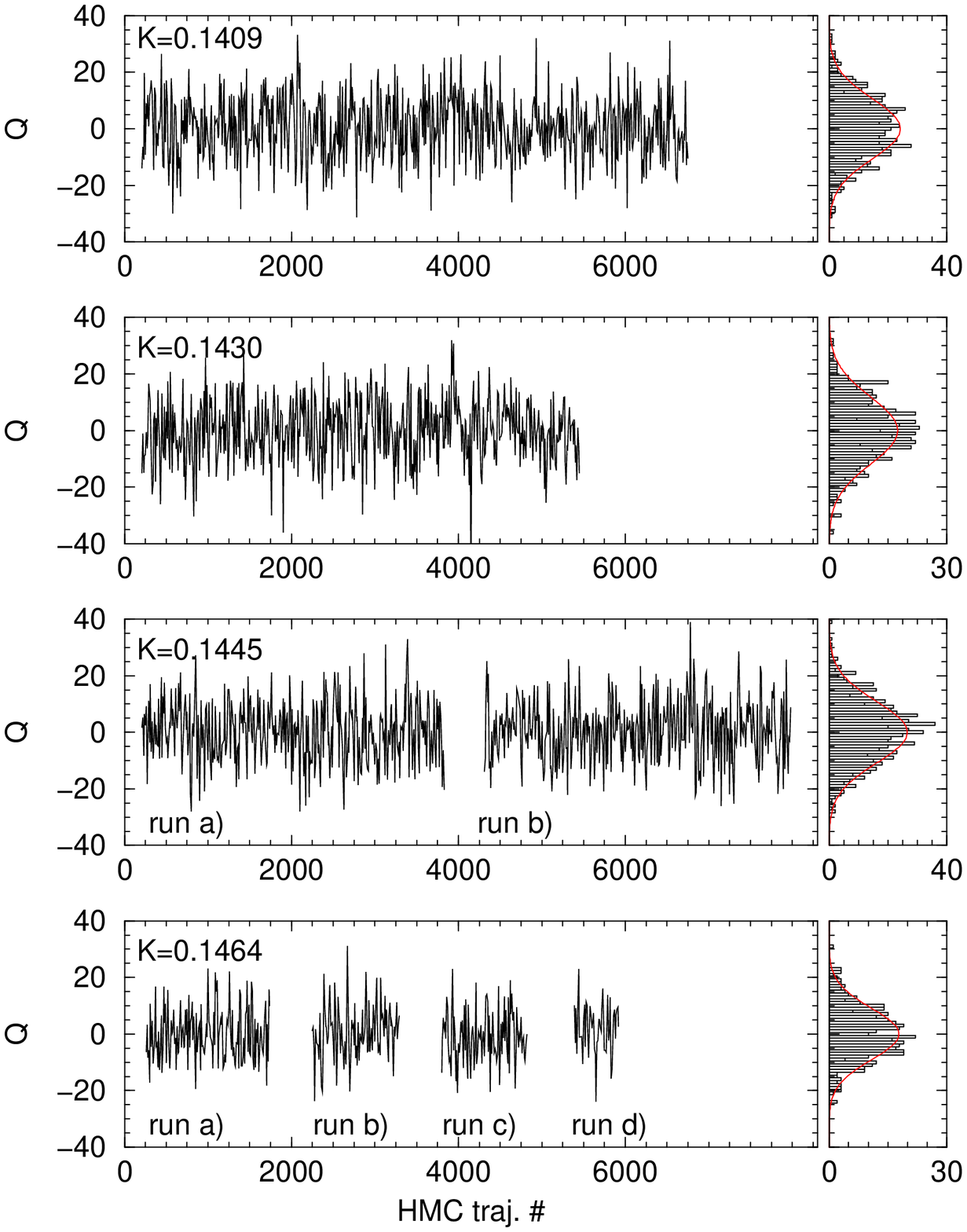}
}
\caption{Time histories and histograms in full QCD at $\beta=1.80$.}
\label{fig:HistB180}
\end{figure*}

\begin{figure*}[p]
\vspace{5mm}
\centerline{
\epsfxsize=10cm \epsfbox{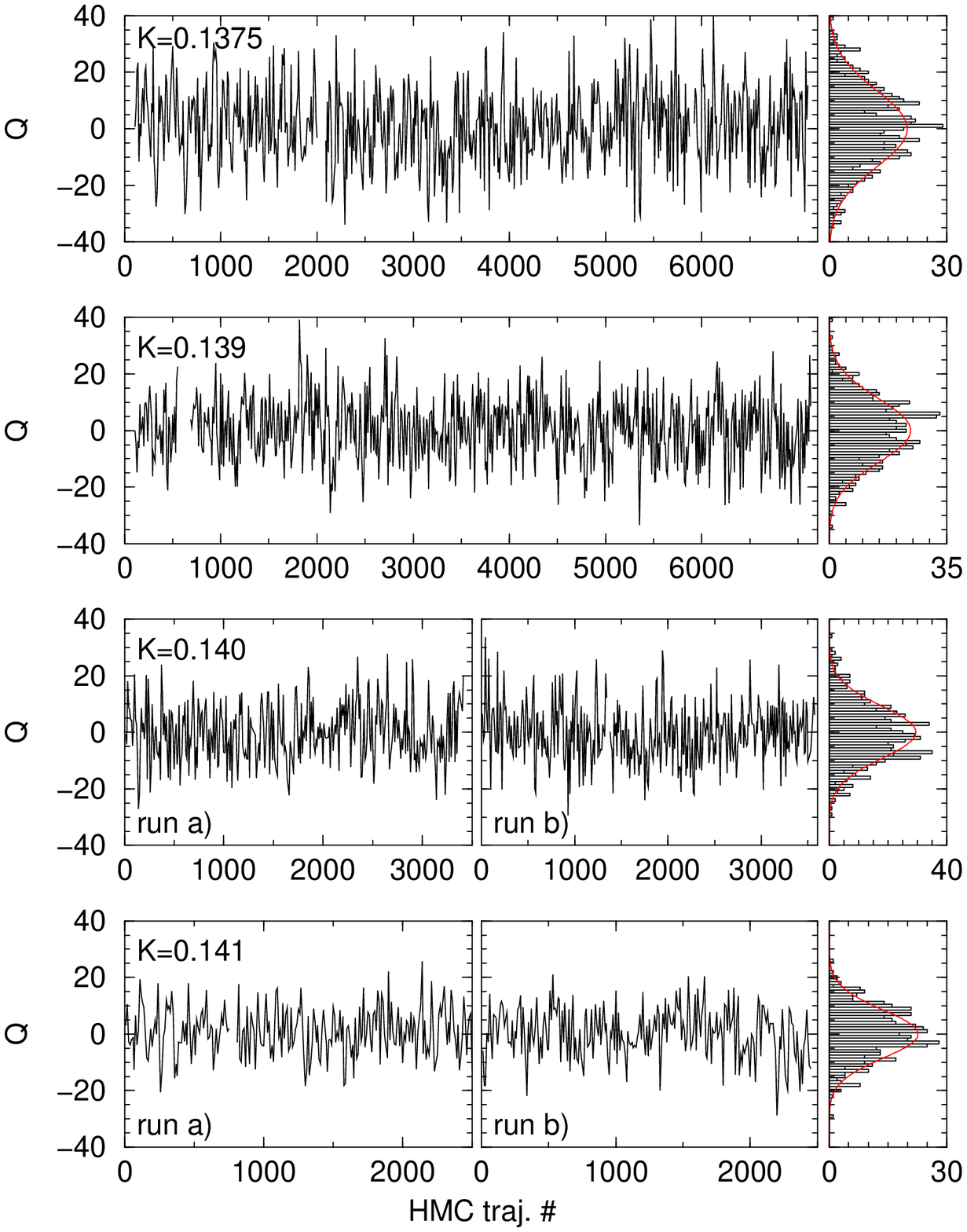}
}
\caption{Time histories and histograms in full QCD at $\beta=1.95$.}
\label{fig:HistB195}
\end{figure*}

\begin{figure*}[p]
\vspace{5mm}
\centerline{
\epsfxsize=10cm \epsfbox{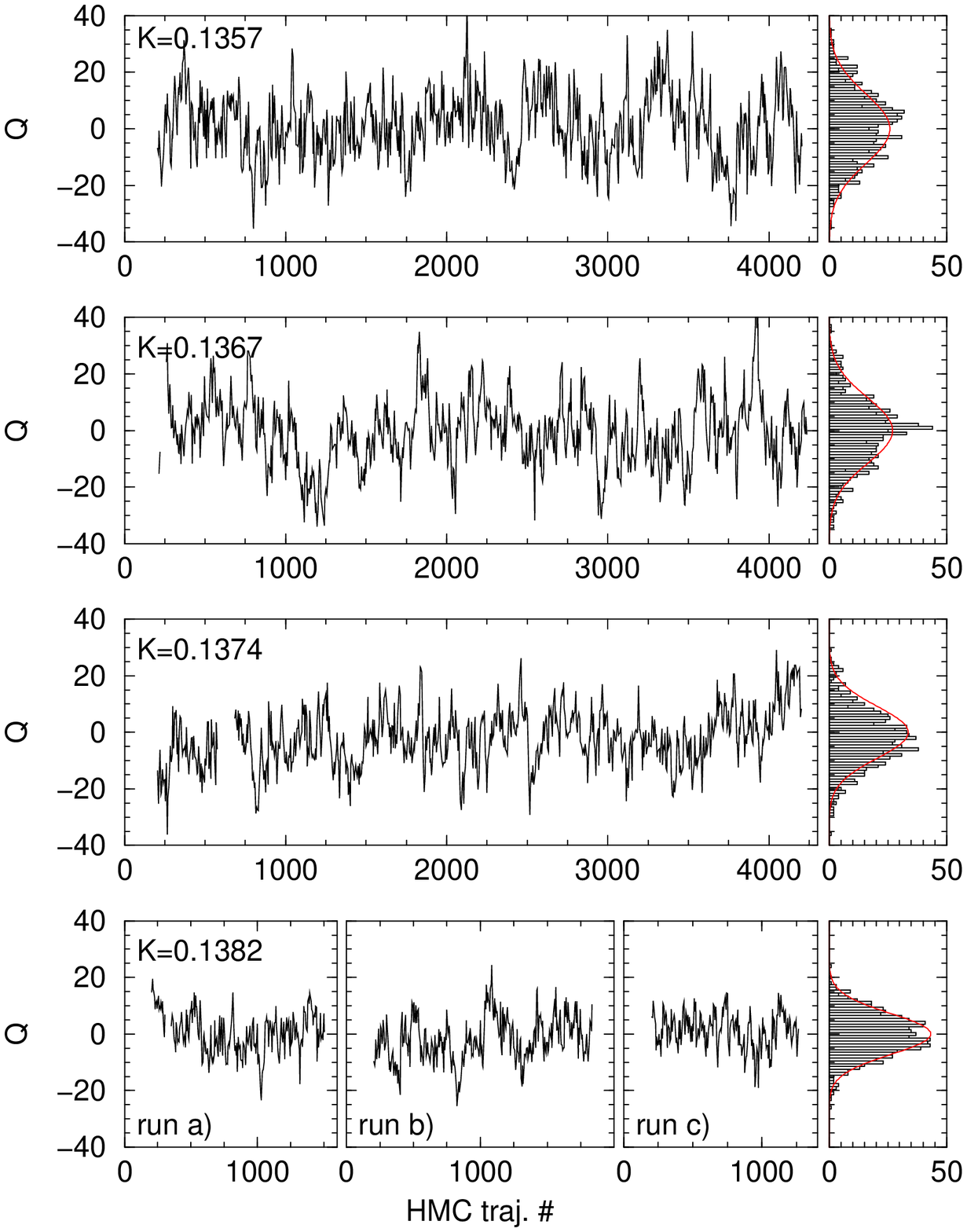}
}
\caption{Time histories and histograms in full QCD at $\beta=2.10$.}
\label{fig:HistB210}
\end{figure*}

In Figs.~\ref{fig:HistB180}, \ref{fig:HistB195} and \ref{fig:HistB210} we
plot time histories of $Q_{\rm imp}$ after 20 cooling steps calculated for
our full QCD runs at all sea quark masses. Autocorrelation times are
visibly small even at the smallest quark masses.  For $\beta=1.80$ and
$\beta=1.95$ the topological charges measured on configurations separated by
10 HMC trajectories are well decorrelated, and hence the integrated
autocorrelation time is smaller than 10 trajectories.  Correspondingly,
errors are independent of the bin size when employing the binning
method. At $\beta=2.10$, where the charge is measured at every fifth HMC
trajectory, we find integrated autocorrelation times of 5--6
configurations, corresponding to 25--30 HMC trajectories. This is
comparable to, but somewhat smaller than, recent results reported for the
Wilson~\cite{SESAM1} or the clover quark action~\cite{Hart}.  For error
estimates throughout this paper we use bins of 10 configurations,
corresponding to 50 HMC trajectories, at $\beta=2.10$ and no binning for
the two other couplings.

A related issue is the ergodicity of HMC simulations. In
Figs.~\ref{fig:HistB180}, \ref{fig:HistB195} and \ref{fig:HistB210} we show
histograms of the topological charge. They are reasonably symmetric around
zero and the distribution can be approximately described by a Gaussian,
also plotted in the figures. Ensemble averages $\langle Q \rangle$, listed
in Table~\ref{tab:suscept}, are consistent with zero or deviate at most
three standard deviations of statistical error at $\beta=2.1$ and
$\kappa=0.1374$. We conclude that topology is well sampled in our runs.


\subsection{Scale determination}

\begin{figure*}[p]
\vspace{5mm}
\centerline{
\epsfxsize=9cm \epsfbox{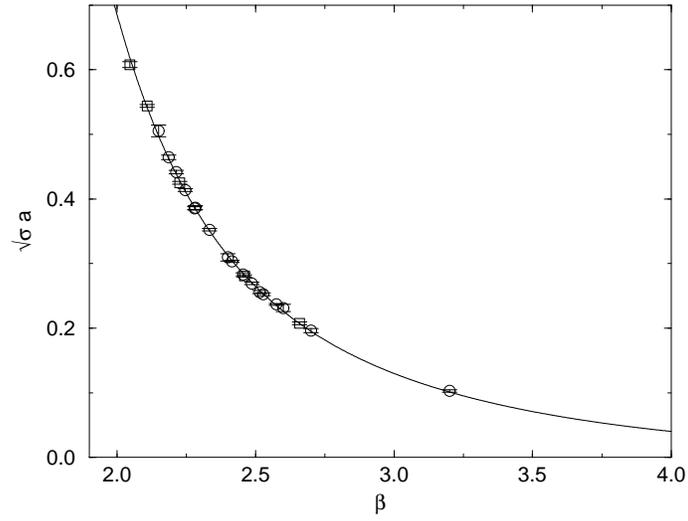}
}
\caption{String tension in pure SU(3) gauge theory as a function of the
gauge coupling. Circles represent data from
Refs.~\protect\cite{full,RGstring,RGstring2} while squares are obtained in
the present work. The solid line represents a fit with
Eq.~(\protect\ref{eq:allton}).}
\label{fig:Sigma}
\end{figure*}

To fix the scale we use the string tension $\sigma$ or the Sommer parameter
$r_0$ \cite{Sommer} of the static quark potential. Full QCD values of $r_0$
have been determined in Ref.~\cite{full} and are reproduced in
Table~\ref{tab:overview}. The analysis of the static quark potential in
pure SU(3) gauge theory of this work parallels the one in Ref.~\cite{full}.
We list $\sigma$ and $r_0$ in Table~\ref{tab:overview-pure}. The dependence
of the dimensionless string tension $\sqrt{\sigma}a$ on the gauge coupling
is shown for pure SU(3) gauge theory in Fig.~\ref{fig:Sigma} together with
previous results of Refs.~\cite{full,RGstring,RGstring2}. Data are
consistent with previous determinations, and extend the domain of results
to smaller values of $\beta$.

We fit the string tension data of Fig.~\ref{fig:Sigma}
using an ansatz proposed by Allton \cite{Allton}, 
\begin{equation}
 \sqrt{\sigma}a 
 = f(\beta) \, \left\{1 + c_2\hat{a}(\beta)^2 
  + c_4\hat{a}(\beta)^4\right\}/c_0, \;\;\;\;
  \hat{a}(\beta) \equiv f(\beta)/f(\beta=2.4),
\label{eq:allton}
\end{equation}
where $f(\beta)$ is the two-loop scaling function of 
SU(3) gauge theory,
\begin{equation}
  f(\beta) = \left(\frac{6b_0}{\beta}\right)^{- b_1/2b_0^2}
                  \exp\left(-\frac{\beta}{12 b_0}\right ), \;\;\;\;
 b_0 = \frac{11}{(4\pi)^2}, \;\; b_1 = \frac{102}{(4\pi)^4}.
\label{eq:allton2}
\end{equation}
We obtain the best fit at 
\begin{equation}
c_{0} = 0.5443(97) ,\;  c_{2} = 0.390(38) ,\;  c_{4} = 0.049(12),
\end{equation}
with good $\chi^{2}/N_{DF} = 19.3/19$. The fit curve plotted in
Fig.~\ref{fig:Sigma} reproduces the data very well.


\section{Topological Susceptibility}
\label{sec:suscept}

\subsection{Pure SU(3) gauge theory}

The topological susceptibility 
\begin{equation}
\chi_t = \frac{\langle Q^2 \rangle}{V}
\end{equation}
in pure SU(3) gauge theory is converted to the dimensionless number $\chi_t
r_0^4$ using measured values of the Sommer scale $r_0$ and is quoted in
Table~\ref{tab:suscept-pure}. Statistical errors of $\langle Q^2 \rangle$
and $r_0$ and the systematic error related to the choice of the number of
cooling steps are added in quadrature.

\begin{figure*}[p]
\vspace{5mm}
\centerline{
              \epsfxsize=7cm \epsfbox{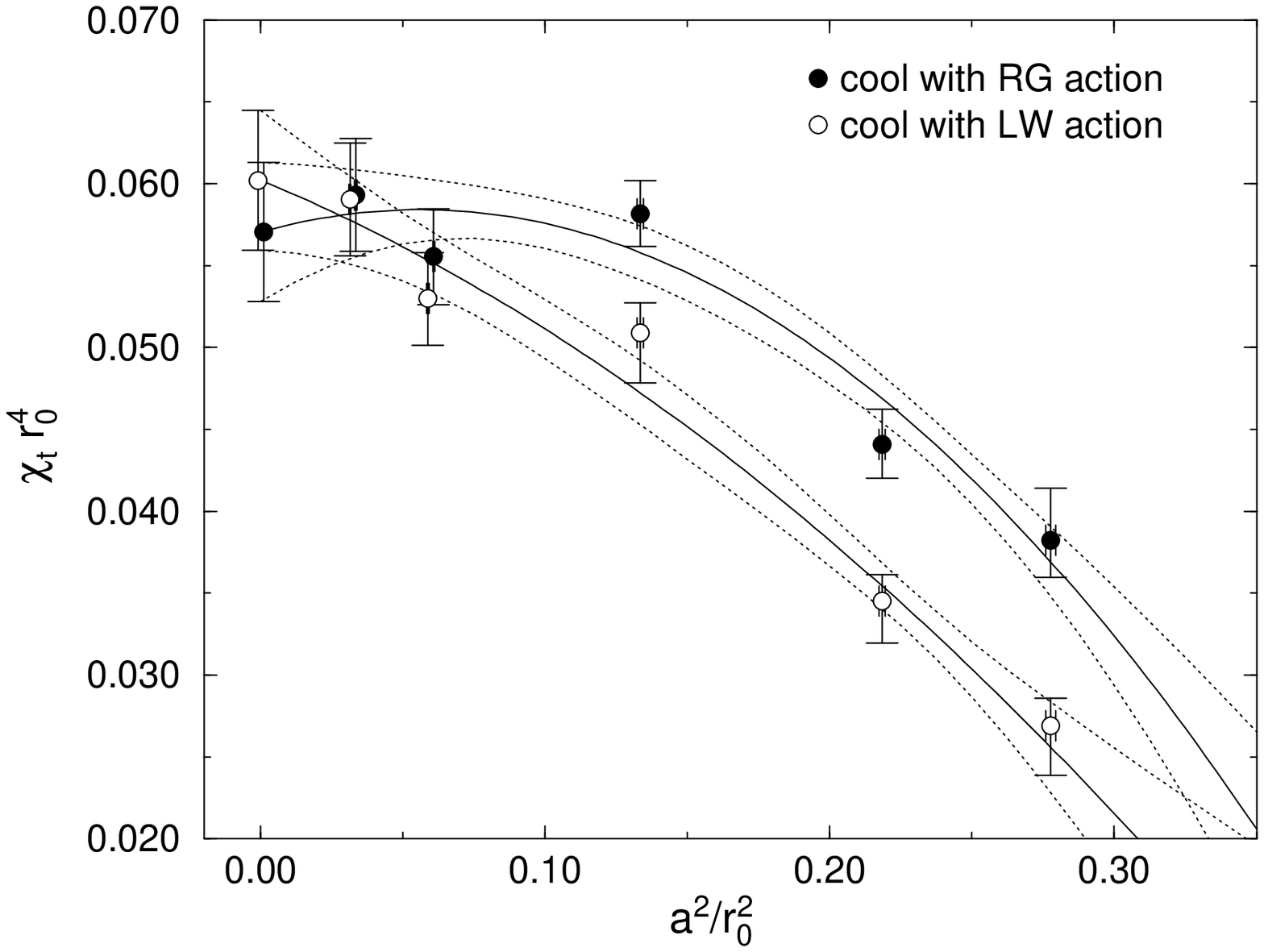}
\hspace{5mm}  \epsfxsize=7cm \epsfbox{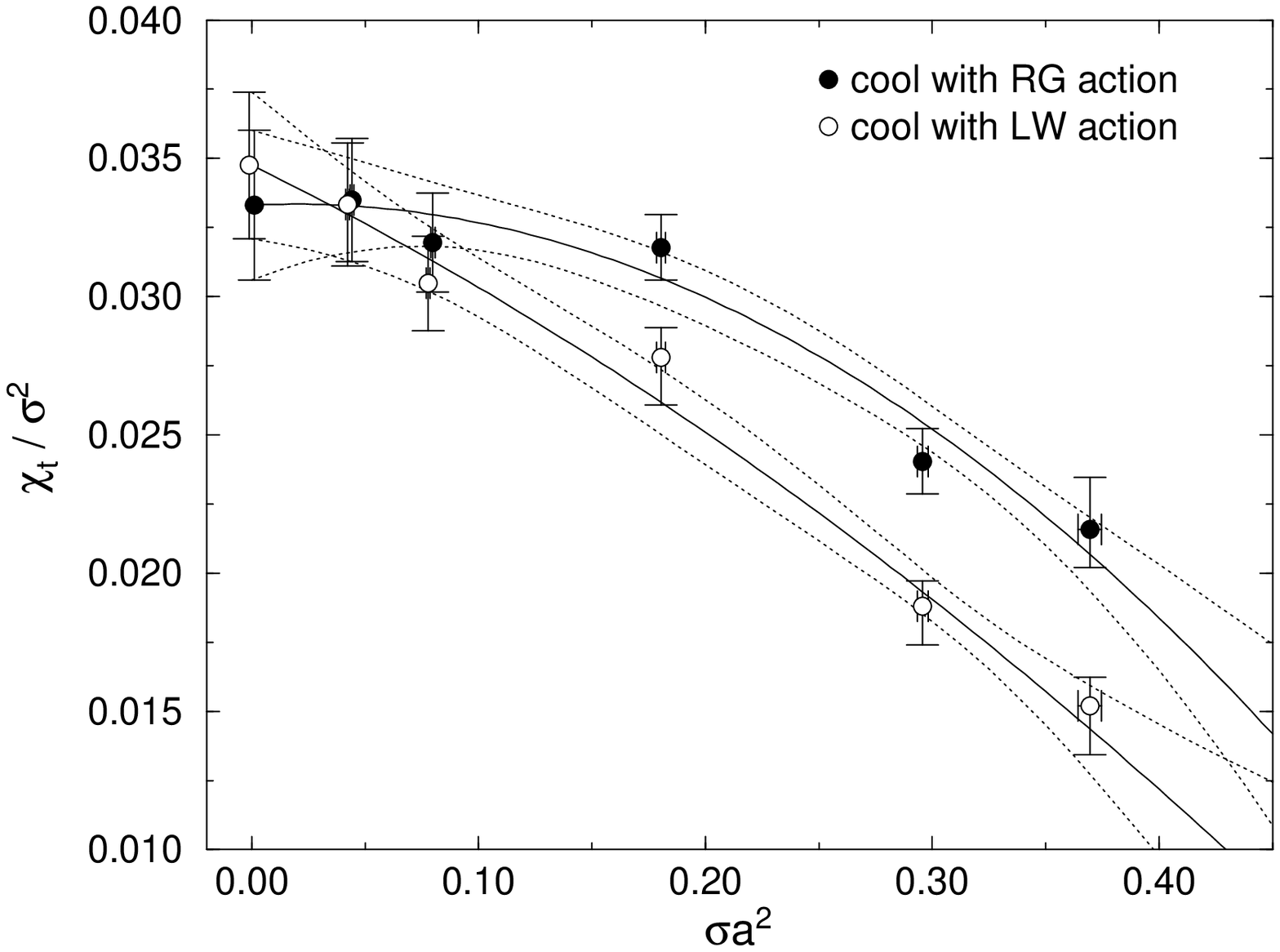}
}
\caption{Continuum extrapolation of the topological susceptibility in pure
SU(3) gauge theory.}
\label{fig:ContExt}
\vspace{-4mm}
\end{figure*}

We plot $\chi_t r_0^4$ as a function of $a^2/r_0^2$ in
Fig.~\ref{fig:ContExt}.  Results obtained with the two cooling actions are
significantly different from each other at coarser lattice spacings. As
expected, they move closer together toward the continuum limit. On the
finest lattice the difference almost vanishes. Since data exhibit a
curvature, we attempt continuum extrapolations including the leading
scaling violation term of $O(a^2)$ and the next higher order term of
$O(a^4)$. We obtain
\begin{equation}
\chi_t r_0^4 = \left\{
\begin{array}{ll}
0.0570(43)  + 0.049(61) a^2/r_0^2 - 0.44(19) a^4/r_0^4 & 
\mbox{cool with $S_{\rm RG}$} \\
0.0602(43)  - 0.072(68) a^2/r_0^2 - 0.19(22) a^4/r_0^4 & 
\mbox{cool with $S_{\rm LW}$}, \\
\end{array}
\right.
\label{eq:chir0SU3Ext}
\end{equation}
with $\chi^{2}/N_{DF} = 2.2$ and 1.4 respectively. Fit curves plotted in
Fig.~\ref{fig:ContExt} follow the data well. In the continuum limit $\chi_t
r_0^4$ obtained with the two cooling actions differ by about one standard
deviation of statistics. 

In Fig.~\ref{fig:ContExt} we also plot $\chi_t$ normalized by the string
tension. Data behave similar to the one normalized by $r_0$. A continuum
extrapolation of the same form as above leads to 
\begin{equation}
\chi_t / \sigma^2 = \left\{
\begin{array}{ll}
0.0333(27)  + 0.004(29) \sigma a^2 - 0.103(67) \sigma^2 a^4 & 
\mbox{cool with $S_{\rm RG}$} \\
0.0347(27)  - 0.040(31) \sigma a^2 - 0.041(76) \sigma^2 a^4 & 
\mbox{cool with $S_{\rm LW}$}, \\
\end{array}
\right.
\label{eq:chisigSU3Ext}
\end{equation}
with $\chi^{2}/N_{DF} = 1.5$ and 0.8 respectively.

To set the scale we use $r_0=0.49(3)$~fm or $\sqrt{\sigma}=440(30)$~MeV
where the errors in parentheses are our estimates of uncertainty of these 
quantities which are not directly measurable in experiments.  
Employing $\chi_tr_0^4$ from cooling with the RG-improved 
action as the central value, we obtain for the
topological susceptibility in pure SU(3) gauge theory,
\begin{equation}
\chi_t^{1/4} = 197(4)(^{+3}_{-0})(^{+0}_{-9})(12) \;\; \mbox{MeV},
\label{eq:chiSU3}
\end{equation}
where the first error is statistical, the second is associated with the
uncertainty from the cooling action, the third reflects the difference from 
using $r_0$ or $\sqrt{\sigma}$ to set the scale, and the last comes from the 
uncertainty in $r_0$.

Our value of $\chi_t^{1/4}$ is in good agreement with recent determinations
by several groups using different
methods~\cite{pisa-SU(3),ImpCool-SU(3),UKQCD-SU3,boulder-SU(3)} as well as
with the Witten-Veneziano relation~\cite{VW}, $\chi_t =
f_\pi^2(m_{\eta'}^2+m_\eta^2-2m_K^2)/2N_f \approx (180\;{\rm MeV})^4$.

\subsection{Full QCD}

\begin{figure*}[p]
\vspace{5mm}
\centerline{
\epsfxsize=8cm \epsfbox{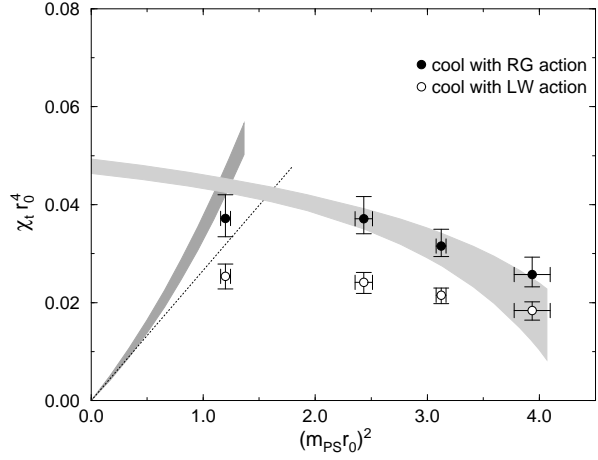}
}
\caption{Topological susceptibility in full QCD at $\beta=1.80$. The light
shaded region indicates the one standard deviation error band for pure
SU(3) gauge theory, cooled with the RG-improved action, 
at corresponding values of
$r_0$. The darker shaded region starting at zero is the one standard
deviation error band of the small mass prediction of
Eq.~(\ref{eq:chiR0smallmass}) evaluated with measured values of $f_{\rm
PS}(m_{\rm PS}^2)$ and $r_0(m_{\rm PS}^2)$ while the dotted line is the
same prediction evaluated with measured values of $f_{\rm PS}$ and $r_0$ at
physical quark masses.}
\label{fig:SuscB180}
\vspace{-4mm}
\end{figure*}

\begin{figure*}[p]
\vspace{5mm}
\centerline{
\epsfxsize=8cm \epsfbox{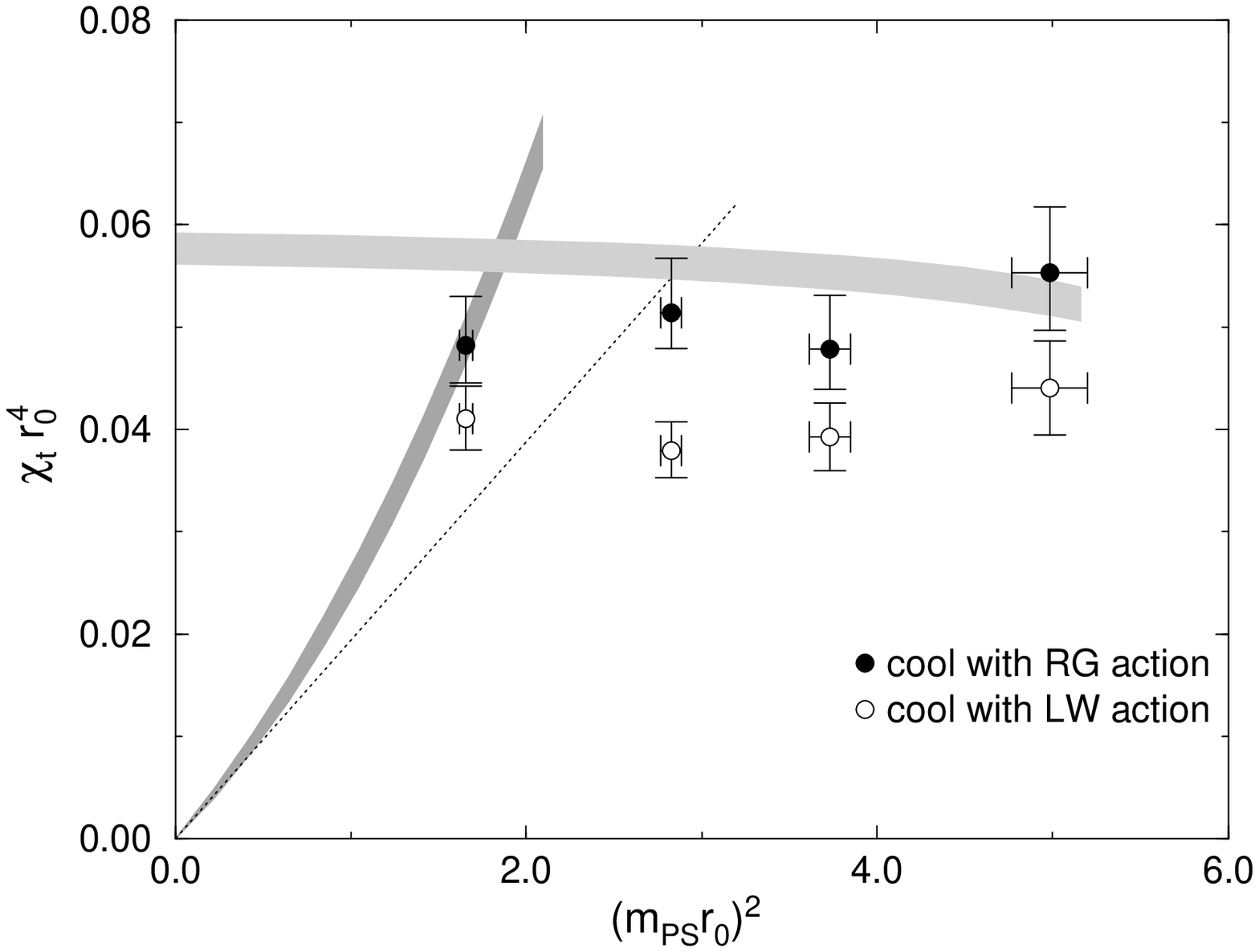}
}
\caption{Topological susceptibility in full QCD at $\beta=1.95$. Symbols
are the same as in Fig.~\ref{fig:SuscB180}.}
\label{fig:SuscB195}
\vspace{-4mm}
\end{figure*}

\begin{figure*}[p]
\vspace{5mm}
\centerline{
\epsfxsize=8cm \epsfbox{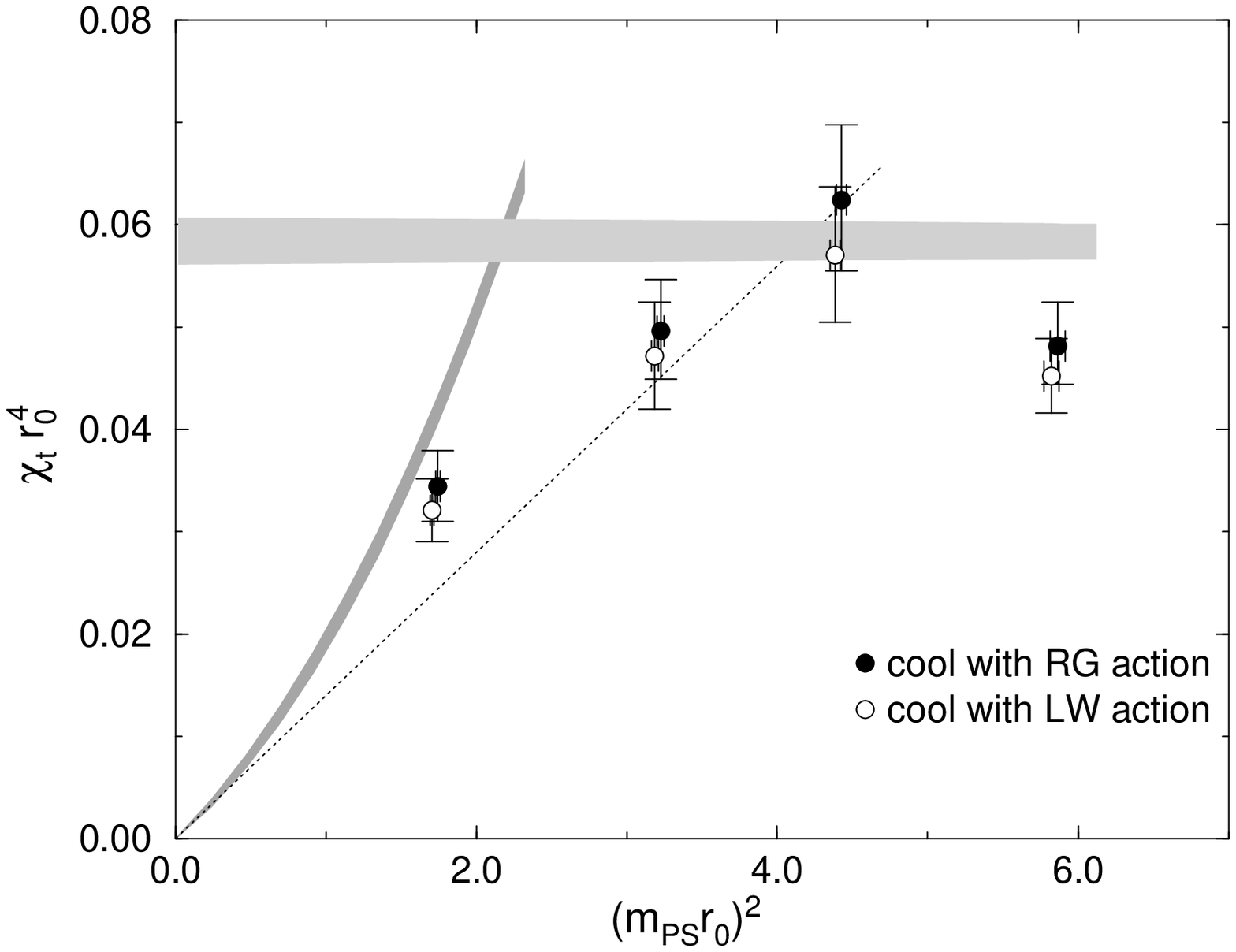}
}
\caption{Topological susceptibility in full QCD at $\beta=2.10$. Symbols
are the same as in Fig.~\ref{fig:SuscB180}.}
\label{fig:SuscB210}
\vspace{-4mm}
\end{figure*}

Topological susceptibilities obtained in full QCD runs and normalized by
$r_0$ measured for the same sea quark mass are collected in
Table~\ref{tab:suscept}. In Figs.~\ref{fig:SuscB180}, \ref{fig:SuscB195}
and \ref{fig:SuscB210} they are plotted as a function of $(m_{\rm PS}
r_0)^2$. As in pure SU(3) gauge theory, data obtained with the two cooling
actions differ from each other at $\beta=1.8$ where the lattice is coarsest
but are consistent with each other within error bars at $\beta=2.1$. The
quark mass dependence is similar between the two cooling actions at all the
$\beta$ values.

For comparison we also plot in Figs.~\ref{fig:SuscB180}, \ref{fig:SuscB195}
and \ref{fig:SuscB210} susceptibilities in pure SU(3) gauge theory obtained
by cooling with the RG-improved action. In full QCD $r_0$ changes together
with the sea quark mass or $m_{\rm PS}$ in Table~\ref{tab:overview}. The
topological susceptibility in pure gauge theory is a decreasing function of
$a^2/r_0^2$, and the value corresponding to full QCD at the same $r_0$ is
therefore not constant when $m_{\rm PS}$ changes. We take this into account
by using the interpolation formula of Eq.~(\ref{eq:chir0SU3Ext}) and the
linear fit of $1/r_0$ as a function of $m_{\rm PS}^2$ in Ref.~\cite{full}
and calculate $\chi_t r_0^4$ at matching values of $r_0$. We arrive at the
one standard deviation error band of the susceptibility in pure gauge
theory plotted as the light shaded area in Figs.~\ref{fig:SuscB180},
\ref{fig:SuscB195} and \ref{fig:SuscB210}. An increasing tendency with
decreasing quark mass is manifest at $\beta=1.8$, whereas at $\beta=2.1$
the shaded error band is very flat.

The topological susceptibility in full QCD is consistent with that of pure
gauge theory at the heaviest quark mass for $\beta=1.8$ and 1.95, but
smaller by two standard deviations for $\beta=2.1$. Values at intermediate
quark masses are consistent or slightly smaller. At the smallest
quark mass the topological susceptibility in full QCD is suppressed
compared to the pure gauge value. The decrease is, however, contained
within 15\% or one to two standard deviations at $\beta=1.8$ and 1.95,
which is marginal. A clearer decrease by 41\%, corresponding to seven
standard deviations, is observed at $\beta=2.1$.

We investigate if the small suppression due to dynamical quarks at the 
two coarser lattice spacings is against expectations by comparing 
the behavior of $\chi_t$ with the prediction of Eq.~(\ref{eq:chismallmass})
for vanishing quark mass. 
Using the Gell-Mann--Oakes--Renner relation~\cite{GOR,GL}
\begin{equation}
\Sigma = \frac{f_\pi^2m_\pi^2}{4m_q},
\label{eq:GOR}
\end{equation}
with $f_\pi$ normalized to be 132~MeV in experiment,
Eq.~(\ref{eq:chismallmass}) can be rewritten as,
\begin{equation}
\chi_t r_0^4 = \frac{(f_{\rm PS} r_0)^2(m_{\rm PS} r_0)^2}{4N_f} 
+ O(m_{\rm PS}^4).
\label{eq:chiR0smallmass}
\end{equation}
In Ref.~\cite{full} pseudoscalar decay constants $f_{\rm PS}$ and Sommer
scale $r_0$ have been determined for all gauge couplings and fitted as
functions of $m_{\rm PS}^2$.  Using the fits $f_{\rm PS}(m_{\rm PS}^2)$ and
$r_0(m_{\rm PS}^2)$ we calculate the one standard deviation error band of
Eq.~(\ref{eq:chiR0smallmass}) and plot it as dark shaded area starting at
zero in Figs.~\ref{fig:SuscB180}, \ref{fig:SuscB195} and
\ref{fig:SuscB210}. We plot the same prediction evaluated with measured
values of $f_{\rm PS}$ and $r_0$ at physical quark masses as dotted line.
Differences between the band and the line are of order $m_{\rm PS}^4$.
Sizable scaling violations in $f_{\rm PS}$ have been observed in
Ref.~\cite{full} with $f_\pi = 195(5)$~MeV ($\beta=1.8$), 157(7)~MeV
($\beta=1.95$) and 131(7)~MeV ($\beta=2.1$) if the scale is 
determined by the $\rho$ meson mass. Correspondingly, the slope of
the prediction Eq.~(\ref{eq:chiR0smallmass}) shows a variation with
$\beta$.
 
The susceptibility in full QCD at the smallest quark mass lie between the
shaded band and the dotted line of
Eq.~(\ref{eq:chiR0smallmass}). Interestingly, the smallest simulated quark
masses at $\beta=1.8$ and 1.95 lie roughly in the region where the small
mass prediction and pure SU(3) gauge theory cross. A stronger suppression 
of the topological susceptibility at $\beta=2.1$, on the other hand, occurs
at a quark mass somewhat below the crossing point. This may be an
indication that the runs at $\beta=2.1$ reach quark masses where a
suppression compared to pure SU(3) gauge theory can be expected. The exact
location of the cross-over region depends, however, on the magnitude of
higher order terms in Eq.~(\ref{eq:chiR0smallmass}) and the lattice value
of $f_{\rm PS}$. Simulations at lighter quark masses will therefore be
helpful to clarify whether the interpretation described here is correct.


\section{Discussion and Conclusions}
\label{sec:conclusions}

We have studied the topological susceptibility as a function of quark mass 
and lattice spacing in two-flavor full QCD using a field theoretic definition 
of the topological charge together with cooling. 

We have shown that an improved charge discretization can be defined which
produces charges close to integers.  The stability of lattice instantons
differs between two actions used for cooling, which leads to different
values of the topological charge at coarse lattice spacings. We have
confirmed that the difference decreases with decreasing lattice spacing and
vanishes in the continuum limit. Our investigation of time histories of the
topological charge in full QCD have shown that autocorrelations are
reasonably short and that our runs are long enough to sample topology well.
These analyses support our belief that systematic errors of the cooling 
method are kept under control, and that our lattice measurements indeed 
reflect topological properties of the QCD vacuum. 

The quark mass dependence of the topological susceptibility $\chi_t r_0^4$
in full QCD is found to be flat or even increase with decreasing quark mass
at $\beta=1.8$ and 1.95, and a clear decrease is only observed at
$\beta=2.1$. A comparison with pure gauge theory at corresponding $r_0$
shows that $\chi_t r_0^4$ in full QCD is consistent with pure gauge theory
at heavier quark masses but suppressed at the lightest quark mass of our
simulation.  At the same time, the susceptibility at the lightest quark 
masses are in agreement with the prediction of the anomalous
Ward-Takahashi identity for U(1) chiral symmetry for small quark masses 
when lattice values for the pseudoscalar decay constant are employed. 
These results suggest that our lightest simulated quark masses lie around 
the transition region where a suppression due to sea quarks is expected to 
set in.

Recently several alternative theoretical explanations have been suggested 
as to why the topological susceptibility in lattice full QCD might appear
less suppressed than expected for small quark masses. 
It has been pointed out~\cite{Durr} that a large enough volume
with $V\Sigma m_q \gg 1$~\cite{LS} is necessary for
Eq.~(\ref{eq:chismallmass}) to be valid. Since we employ a large lattice
size of $La\approx 2.5$~fm and quark masses with $m_q \simgt 40$~MeV this
condition is always fulfilled.  It has also been argued that subtleties
exist in the flavor singlet lattice Ward-Takahashi identities when the
Wilson or clover fermion action is employed so that counter-terms are 
needed for the correct chiral behavior of the topological
susceptibility~\cite{Rossi}.  Simulations at lighter sea quark masses and
further theoretical analyses are needed to examine whether such an
explanation is required for understanding the quark mass dependence of
$\chi_t$ in full QCD.


\acknowledgements

This work was supported in part by Grants-in-Aid of the Ministry of
Education (Nos. 09304029, 10640246, 10640248, 10740107, 11640250, 11640294,
11740162, 12014202, 12304011, 12640253, 12740133, 13640260). AAK and TM were
supported by the JSPS Research for the Future Program (No. JSPS-RFTF
97P01102). SE, KN and HPS were JSPS Research Fellows.



\newpage


\begin{table}
\caption{Overview of full QCD simulations.  The lattice spacing $a$ is
fixed by the vector meson mass at the physical quark mass and $\protect
M_\rho=768.4$~MeV.}
\label{tab:overview}
\begin{tabular}{cccccccccccc}
$\beta$ & $L^3\times T$ & $c_{SW}$ & $a$ [fm] & $La$ [fm] & $\kappa$ &
 $m_{\rm PS}a$ & $m_{\rm PS}/m_{\rm V}$ & $r_0/a$ & $N_{\rm Meas}$ & 
$N_{\rm Skip}$ & $N_{\rm Bin}$ \\
\tableline 
1.80 & $12^3{\times}24$ & $1.60$ & 0.2150(22) & 2.580(26)  &
       0.1409  &  1.15601(61) & 0.807(1) & 1.716(35) & 650  & 10 & 1 \\
  &&&&& 0.1430  &  0.98267(89) & 0.753(1) & 1.799(13) & 522  & 10 & 1 \\
  &&&&& 0.1445  &  0.82249(82) & 0.694(2) & 1.897(30) & 729  & 10 & 1 \\
  &&&&& 0.1464  &  0.5306(17)  & 0.547(4) & 2.064(38) & 409  & 10 & 1 \\ 
1.95 & $16^3{\times}32$ & $1.53$ & 0.1555(17) & 2.489(27) &
       0.1375  &  0.89400(52) & 0.804(1) & 2.497(54) & 681  & 10 & 1 \\
  &&&&& 0.1390  &  0.72857(68) & 0.752(1) & 2.651(42) & 690  & 10 & 1 \\
  &&&&& 0.1400  &  0.59580(69) & 0.690(1) & 2.821(29) & 689  & 10 & 1 \\
  &&&&& 0.1410  &  0.42700(98) & 0.582(3) & 3.014(33) & 488  & 10 & 1 \\ 
2.10 & $24^3{\times}48$ & $1.47$ & 0.1076(13) & 2.583(31) &
       0.1357  &  0.63010(61) & 0.806(1) & 3.843(16) & 800  & 5  & 10 \\
  &&&&& 0.1367  &  0.51671(67) & 0.755(2) & 4.072(15) & 788  & 5  & 10 \\
  &&&&& 0.1374  &  0.42401(46) & 0.691(3) & 4.236(14) & 779  & 5  & 10 \\
  &&&&& 0.1382  &  0.29459(85) & 0.576(3) & 4.485(12) & 789  & 5  & 10 \\ 
\end{tabular}
\end{table}

\begin{table}
\caption{Overview of pure SU(3) simulations. The lattice spacing $a$ is
determined using $\sqrt{\sigma}=440$~MeV. Numbers in
brackets for $N_{\rm Conf}$ indicate the number of configurations used for 
potential measurement.} 
\label{tab:overview-pure}
\begin{tabular}{ccccccc}
$\beta$ & $L^3\times T$ & $a$ [fm] & $La$ [fm] & $\sigma a^2$ & 
$r_0/a$ & $N_{\rm Conf}$ \\
\tableline
2.047 &  $8^3{\times}8$  & 0.2726(19) & 2.181(15) & 0.3695(52)  & 1.8978(59) &  500  \\  
2.110 &  $8^3{\times}8$  & 0.2439(10) & 1.951(8)  & 0.2958(24)  & 2.1399(53) & 1000  \\  
2.227 &  $8^3{\times}8$  & 0.1905(10) & 1.524(8)  & 0.1805(19)  & 2.738(11)  & 2000  \\  
2.461 & $12^3{\times}12$ & 0.1259(7)  & 1.511(9)  & 0.07885(90) & 4.089(14) &  900  \\  
2.659 & $16^3{\times}16$ & 0.0931(9)  & 1.489(14) & 0.04311(84) & 5.556(30)  &  700(495)  \\  
\end{tabular}
\end{table}

\begin{table}
\caption{Ratio between center of peak of the topological charge
distribution and integer charge after 10/20/50 cooling steps with the 
RG-improved action. 
At missing numbers no clearly separated peak structure could be
identified.}
\label{tab:DevInt}
\begin{tabular}{ccc}
$\beta$ & standard Q & improved Q   \\
\tableline
2.047 &  ---/0.77/0.85 &  ---/0.94/0.97  \\  
2.110 &  ---/0.80/0.87 & 0.89/0.94/0.98  \\
2.227 & 0.78/0.83/0.88 & 0.94/0.96/0.98  \\
2.461 & 0.85/0.89/0.92 & 0.97/0.98/0.99  \\
2.659 & 0.89/0.92/0.94 & 0.98/0.99/0.995  \\ 
\end{tabular}
\end{table}

\begin{table}
\caption{Correlation coefficient $r$ between $Q_{\rm imp}$ obtained after 10,
20 or 50 cooling steps with the RG-improved or the LW action.}
\label{tab:corrRGLW}
\begin{tabular}{cccc}
$\beta$ & 10 steps & 20 steps & 50 steps \\
\tableline
2.047 & 0.90(1)   & 0.86(1)  & 0.84(1)   \\  
2.110 & 0.923(5)  & 0.886(7) & 0.871(8)  \\
2.227 & 0.961(2)  & 0.942(3) & 0.931(3)  \\
2.461 & 0.991(1)  & 0.986(2) & 0.982(3)  \\
2.659 & 0.9982(5) & 0.9978(7)& 0.9970(9) \\ 
\end{tabular}
\end{table}

\begin{table}
\caption{Topological susceptibility in pure SU(3) gauge theory. 
For $\langle Q^2 \rangle$ the first error is statistical, and the second
error is an estimate of systematic error related
to the choice of the number of cooling steps. For $\chi_t r_0^4$ the two 
errors and the statistical error of $r_0$ are added in quadrature.}
\label{tab:suscept-pure}
\begin{tabular}{cccc|ccc}
  &
\multicolumn{3}{c|}{cool with RG-improved action:} &
\multicolumn{3}{c}{cool with LW action:} \\
$\beta$ & $\langle Q \rangle$ & $\langle Q^2 \rangle$ & 
$\chi_t r_0^4$ & $\langle Q \rangle$ & $\langle Q^2 \rangle$ & 
$\chi_t r_0^4$ \\
\tableline 
2.047 &    0.05(15)  & 12.07(69)(+72)   & $0.0382^{+32}_{-22}$
      &    0.00(13)  &  8.50(52)($-$80) & $0.0269^{+17}_{-30}$ \STRUT \\[0.4ex]
2.110 &    0.121(93) &  8.61(39)(+13)   & $0.0441^{+21}_{-20}$
      &    0.054(82) &  6.74(31)($-$38) & $0.0345^{+16}_{-25}$  \\[0.4ex]
2.227 & $-$0.043(46) &  4.24(13)(0)     & $0.0582^{+20}_{-20}$
      & $-$0.042(43) &  3.71(12)($-$18) & $0.0509^{+18}_{-31}$  \\[0.4ex]
2.461 &    0.139(68) &  4.12(21)(0)     & $0.0555^{+29}_{-29}$
      &    0.123(66) &  3.93(20)($-$4)  & $0.0530^{+28}_{-28}$  \\[0.4ex]
2.659 &    0.067(76) &  4.08(22)(0)     & $0.0593^{+34}_{-34}$
      &    0.073(76) &  4.06(22)($-$1)  & $0.0590^{+34}_{-34}$  \\[0.4ex]
\end{tabular}
\end{table}

\begin{table}
\caption{Topological susceptibility in full QCD. Meaning of errors are the 
same as in Table \ref{tab:suscept-pure}.} 
\label{tab:suscept}
\begin{tabular}{ccccc|ccc}
  &  &
\multicolumn{3}{c|}{cool with RG-improved action:} &
\multicolumn{3}{c}{cool with LW action:} \\
$\beta$ & $\kappa$ & $\langle Q \rangle$ & $\langle Q^2 \rangle$ & 
$\chi_t r_0^4$ & $\langle Q \rangle$ & $\langle Q^2 \rangle$ & 
$\chi_t r_0^4$ \\
\tableline 
1.80 
 & 0.1409 &    0.41(44) & 123.4(6.5)(+11.9)  & $0.0258^{+35}_{-25}$
          & $-0$.04(37) &  87.9(4.6)($-3$.3) & $0.0184^{+18}_{-19}$ 
                                                              \STRUT \\[0.4ex]
 & 0.1430 &    0.10(49) & 125.1(7.8)(+10.5)  & $0.0316^{+34}_{-22}$
          &    0.20(40) &  85.2(5.5)($-3$.2) & $0.0215^{+15}_{-17}$  \\[0.4ex]
 & 0.1445 &    0.46(40) & 119.3(6.0)(+10.7)  & $0.0371^{+45}_{-30}$
          &    0.53(33) &  77.7(4.2)($-3$.0) & $0.0241^{+20}_{-22}$  \\[0.4ex]
 & 0.1464 &    0.18(46) &  85.0(5.9)(+6.8)   & $0.0372^{+48}_{-38}$
          &    0.20(38) &  58.1(3.9)($-0$.7) & $0.0254^{+25}_{-25}$  \\[0.4ex]
1.95 
 & 0.1375 &    1.09(52) & 186.4(9.7)(+10.7)  & $0.0553^{+64}_{-56}$
          &    1.35(46) & 148.5(8.4)(+1.6)   & $0.0440^{+46}_{-46}$  \\[0.4ex]
 & 0.1390 &    0.42(43) & 127.0(6.5)(+9.3)   & $0.0478^{+52}_{-39}$
          &    0.27(39) & 104.2(5.7)($-0$.6) & $0.0393^{+33}_{-33}$  \\[0.4ex]
 & 0.1400 & $-0$.33(39) & 106.3(5.7)(+8.3)   & $0.0514^{+53}_{-35}$
          & $-0$.02(34) &  78.4(4.4)(+1.9)   & $0.0379^{+28}_{-27}$  \\[0.4ex]
 & 0.1410 &    0.61(40) &  76.5(4.7)(+5.0)   & $0.0482^{+48}_{-36}$
          &    0.27(37) &  65.2(3.9)(+1.4)   & $0.0411^{+32}_{-31}$  \\[0.4ex]
2.10 
 & 0.1357 &    0.96(88) & 146.4(11.0)(+6.8)  & $0.0481^{+43}_{-37}$  
          &    1.17(88) & 137.4(10.6)(+2.6)  & $0.0452^{+37}_{-36}$  \\[0.4ex]
 & 0.1367 & $-0$.5(1.0) & 150.7(16.6)(+5.7)  & $0.0624^{+73}_{-69}$
          & $-0$.6(1.0) & 137.6(15.6)(+3.4)  & $0.0570^{+67}_{-65}$  \\[0.4ex]
 & 0.1374 & $-2$.52(81) & 102.2(9.7)(+3.3)   & $0.0496^{+50}_{-47}$
          & $-2$.61(82) &  98.0(9.7)(+0.9)   & $0.0472^{+52}_{-52}$  \\[0.4ex]
 & 0.1382 & $-0$.29(63) &  56.5(5.6)(+0.9)   & $0.0344^{+35}_{-34}$
          & $-0$.33(61) &  52.6(5.0)(+0.5)   & $0.0321^{+31}_{-31}$  \\[0.4ex]
\end{tabular}
\end{table}



\newpage

\newpage

\newpage

\newpage

\newpage

\end{document}